\documentclass[fleqn,10pt]{wlscirep}
\usepackage[utf8]{inputenc}
\usepackage[T1]{fontenc}
\usepackage{lineno}
\usepackage{lettrine}
\usepackage{tikz}

\title{Front-induced transitions control THz waves}

\author[1,+]{Aidan W. Schiff-Kearn}
\author[1,2,+]{Lauren Gingras}
\author[1]{Simon Bernier}
\author[1]{Nima Chamanara}
\author[1]{Kartiek Agarwal}
\author[2]{Jean-Michel M\'enard}
\author[1]{David G. Cooke*}
\affil[1]{Department of Physics and Centre for the Physics of Materials, McGill University, Montreal, H3A 2T8 Canada}
\affil[2]{Department of Physics, University of Ottawa, Ottawa, K1N 6N5 Canada}
\affil[*]{Corresponding author: cooke@physics.mcgill.ca}
\affil[+]{these authors contributed equally to this work}

\begin{abstract} %commented out line 169 in wlscirep.cls to kill "ABSTRACT"
\end{abstract}

\begin{document}
\flushbottom
\maketitle
\noindent\textbf{Relativistically moving dielectric perturbations can be used to manipulate light in new and exciting ways beyond the capabilities of traditional nonlinear optics. Adiabatic interaction with the moving front modulates the wave simultaneously in both space and time, and manifests a front-induced transition in both wave vector and frequency yielding exotic effects including non-reciprocity and time-reversal~\cite{GaafarNatPhoton2019,GaafarAPLPhoton2020,BiancalanaPRE2007,GeltnerAPL2002,LampePhysFluids1978,SemenovaRadioPhys1967,CalozPhysRevApp2018}. Here, we introduce a technique called SLIPSTREAM, Spacetime Light-Induced Photonic STRucturEs for Advanced Manipulation. The technique is based on the creation of relativistic fronts in a semiconductor-filled planar waveguide by photoexcitation of mobile charge carriers. Here we demonstrate the capabilities of SLIPSTREAM for novel manipulation of THz light pulses through relativistic front-induced transitions. In the sub-luminal front velocity regime, we generate temporally stretched THz waveforms, with a quasi-static field lasting for several picoseconds tunable with the front interaction distance. In the super-luminal regime, the carrier front outpaces the THz pulse and a time-reversal operation is performed via a front-induced intra-band transition. We anticipate our platform will be a versatile tool for future applications in the THz spectral band requiring direct and advanced control of light at the sub-cycle level.}

\indent Light in the terahertz (THz) band, already immensely important for spectroscopy and imaging applications \cite{JepsenLPR2011}, has been touted as a frontier for next generation high-speed wireless communications~\cite{NagatsumaOptExp2013, MittlemanJAP2017, SenguptaNatElectron2018}. As with other spectral bands, the advancement of future THz technologies will be done in lockstep with the development of new methods for controlling the spatial and temporal properties of THz light. While significant advances have been made on controlling the spatial evolution of THz waves with metasurfaces and wavefront engineering~\cite{ChanAPL2009,YuJOSAB2010, ShabanpourSciRep2020, WattsNatPhoton2014}, methods for actively controlling the temporal properties of THz light on the level of the electric field cycle are typically only possible through self-induced temporal phase modulation using intense THz pulses~\cite{GanichevBook,HegmannPRB2009, HoffmannAPL2009,LiuNature2012,TurchinovichPRB2012,SharmaOptExp2012,AlNaibPRB2013, RaabNatComm2020,HafezAdvOptMater2020}. Only recently an arbitrary THz pulse shaper was introduced, employing a hybrid digital wave synthesis technique using light-induced structures in a waveguide~\cite{Gingras17}. Such planar waveguide approaches have the benefit of being amenable to chip-scale integration~\cite{MaNatComm2017}.\\
\indent Alternative approaches to nonlinear optics, however, draw on the concept of time-varying linear media introduced first by Morgenthaler~\cite{Morgenthaler1958,BiancalanaPRE2007,CalozPhysRevApp2018}. These ideas were tested in experiments involving plasma physics via the rapid ionization of gases \cite{YablonovichPRL1973} or sudden carrier injection through optical excitation in semiconductors~\cite{NishidaAPL2012}. Temporal modulation is attractive for wavelength conversion as it does not require nonlinear media or strong field interactions. Subsequently, several studies have emerged, focused on temporally modulated photonic crystals~\cite{TanakaNatMater2007,NotomiRepProgPhys2010} and cavities~\cite{XuNatPhys2007,SatoNatPhoton2012} for wavelength conversion applications. The non-adiabatic modulation of meta-atoms was recently demonstrated as a means to achieve wavelength conversion in the THz band~\cite{LeeNatPhoton2018}. In addition, sub-cycle slicing using intense optical pulses illuminating semiconductor surfaces is also capable of pulse narrowing and spectral expansion in the THz band~\cite{MayerNJPhys2014,ShalabyAPL2015}.\\
\indent Simultaneous control over the wavevector and frequency of light through combined spatial and temporal modulation can produce a variety of exotic photonic operations~\cite{CalozIEEETrans2020a,CalozIEEETrans2020b, GaafarNatPhoton2019}. These spacetime, front-induced transitions can be used to time-reverse a light pulse,\cite{GaafarAPLPhoton2020,YanikPRL2004,Chumak2010}, for example, which has important applications in dispersion compensation~\cite{YarivOptLett1979} and imaging through random scattering media~\cite{MoskNatPhoton2012}. THz interactions with moving fronts were first investigated in pioneering experiments using photoexcitation in a semiconductor to create an overdense, reflective plasma interaction that is counter-propagating with the incoming THz wave. Considerable spectral broadening was observed via relativistic Doppler shifts~\cite{ThomsonPRB2013,Roskos14,Kohno20}. Here we introduce SLIPSTREAM for exploring front-induced transitions in the THz band and use it to demonstrate two unconventional photonic operations: the temporal stretching and time-reversal of THz pulses.\\
\begin{figure*}[th!]
\centering
\includegraphics[width=\textwidth]{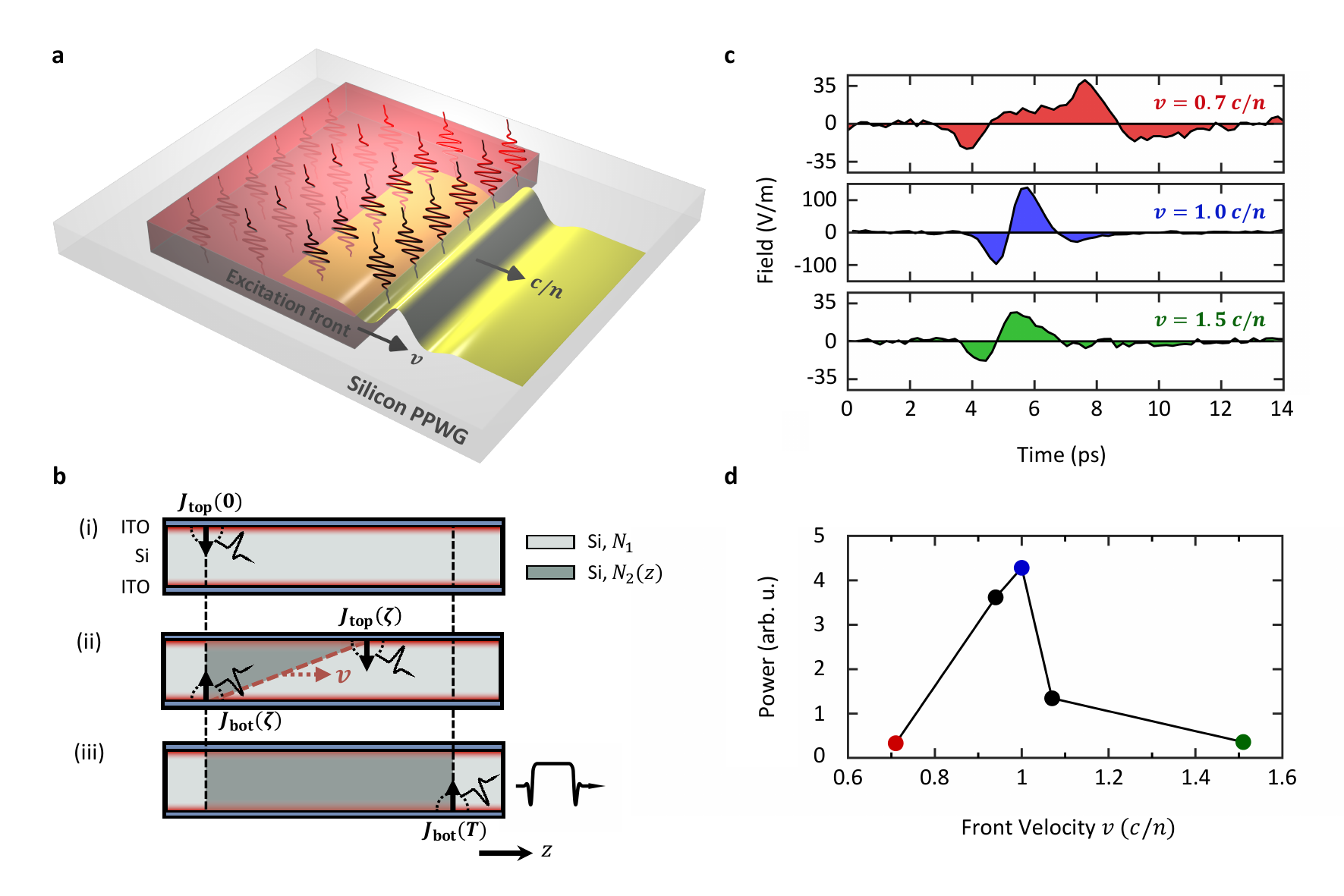}
\caption{\textbf{a}, Schematic of the SLIPSTREAM platform, generating and manipulating THz pulses using the injection of a NIR pulse with a tilted intensity profile into a Si-filled PPWG. \textbf{b}, Symbolic construction of a THz plateau pulse by the space-time-delayed, ultrafast switching of static Schottky currents built into the top and bottom interfaces, shown in successive snapshots in time. \textbf{c}, Measured THz transients typical of the sub-luminal, luminal and super-luminal regimes of front velocity for a maximal induced photocarrier density of about $5\times10^{16}$ cm$^3$. \textbf{d}, Integrated spectral power, peaking for the luminal case due to synchronicity with the emission, maximizing interaction with the moving front.}
\label{fig:figure1}
\end{figure*}
\indent The SLIPSTREAM technique is based on the same compact platform for spatio-temporal manipulation we have previously introduced, capable of performing a range of photonic operations including pulse steering~\cite{CookeAPL2009}, mode control~\cite{Gingras14}, spectral shaping\cite{Gingras16}, arbitrary pulse shaping~\cite{Gingras17} and phase control~\cite{Gingras18}. As shown schematically in Fig.~\ref{fig:figure1}a, a parallel plate waveguide (PPWG) supporting dispersionless, low loss, TEM THz propagation is formed by coating both sides of a $d = 50$-$\mu$m-thick high resistivity float zone silicon ($\rho>10,000 $ $\Omega$-cm) slab with optically transparent and conducting indium tin oxide (sheet resistance of 1~$\Omega$/sq). The transparency of the ITO coatings allows optical excitation of the silicon through the top plate with a near-infrared (1035 nm, 190 fs, 30~$\mu$J) pump pulse derived from a Yb:KGW amplified femtosecond laser system (Light Conversion, PHAROS). This excitation is close to the indirect band gap of Si, with pump penetration depths exceeding 100 $\mu$m, ensuring a homogeneous excitation across the Si slab. The pump pulse has its pulse front tilted using a diffraction grating, producing a linear gradient in the pump arrival time along the waveguide propagation axis. In this manner, as the pump pulse front illuminates the silicon, a moving front of photoinduced carrier density is created within the PPWG. A portion of the NIR beam is split off to a CMOS camera to record the pump intensity spatial profile.\\
\indent The photoexcitation of a moving front results in the emission of a phase-locked THz pulse within the waveguide that propagates in the TEM mode. Since no electrical bias is applied and the plates are grounded, the sole source of THz emission is due to transient currents created by built-in Schottky fields, oppositely oriented at the metal-semiconductor interfaces, upon arrival of the pump pulse at the top and bottom waveguide plates, as in Fig.~\ref{fig:figure1}b. Photoinjecting free carriers within the band bending regions at these interfaces leads to the transient currents $J_\text{top}(t,z)$ and $J_\text{bot}(t-\zeta,z)$ directed normal to the waveguide plates and temporally separated by the transit time of the NIR pump pulse $\zeta$. These delayed and transient currents emit THz light in the same manner as an Auston switch and similar to a continuous version of a segmented dc-to-ac radiation converter~\cite{MoriPhysRevLett1995,HigashiguchiPRL2000,Hashim01,OhataSPIE07,BaeAPL2009,KostinNJP2015,Otsuka17,FukudaJJAP2020}. Once coupled out of the PPWG, the THz emission is detected by free-space electro-optic sampling in a 5-mm-thick (110) GaP crystal. \\
\begin{figure}[!ht]
\centering
\includegraphics[width=0.5\textwidth]{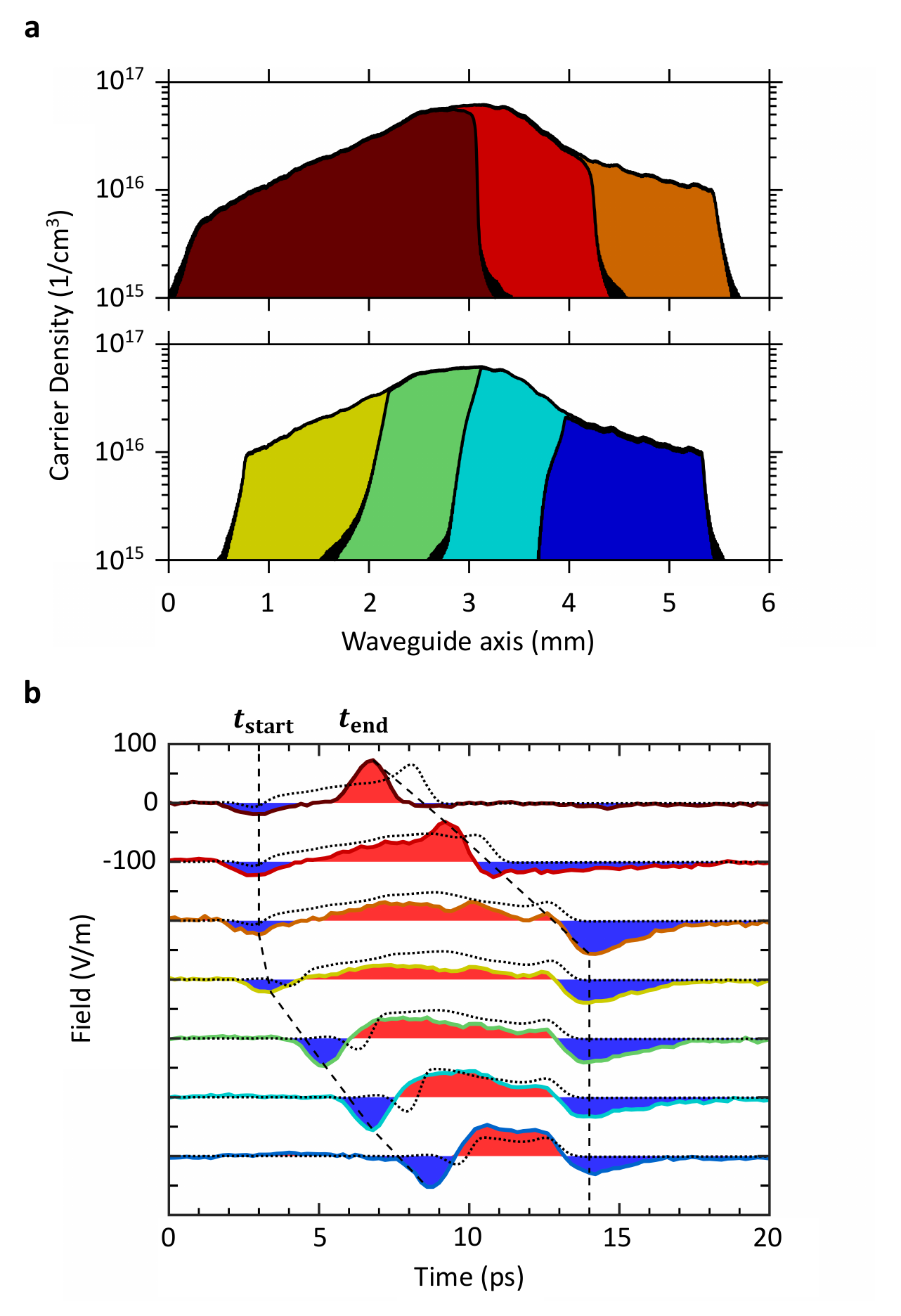}
\caption{\textbf{a}, Carrier density profiles in the waveguide for different beam clippings of a sub-luminal front at $v_f=0.86$ $c/n_\text{Si}$, varying either its termination (top) or start position (bottom). \textbf{b}, Corresponding measured (solid and filled) and simulated (dashed) THz plateau pulses.}
\label{fig:figure2}
\end{figure}
\indent The emitted THz electric field transients, shown in Fig.~\ref{fig:figure1}c, depend critically on the pulse front velocity, $v_f$, with respect to the THz TEM mode phase velocity $c/n_\text{Si}$ with refractive index $n_\text{Si} = 3.418$~\cite{Grisch90}. The front velocity is continuously tunable from the sub-luminal to super-luminal regime with the femtosecond pump pulse front tilt angle. In the sub-luminal regime, $v_f < c/n_\text{Si}$, a THz plateau pulse is created with a stretch of quasi-dc electric field bookended by unipolar structure, shown in the top panel of Fig.~\ref{fig:figure1}c. The emitted THz waves escape the excitation front which produces a superposition of linearly delayed single cycle pulses along the propagation axis. The result is a temporally stretched THz pulse whose duration is tunable with front velocity and total propagation distance. At the luminal condition $v_f = c/n_\text{Si}$, the emitted THz pulse is of single cycle duration, as all emission points add in phase along the entire propagation axis. This produces the maximum amplitude THz emission as shown in the integrated spectral power in Fig.~\ref{fig:figure1}d. Tuning the front to the super-luminal regime $v_f = 1.5$ $c/n_\text{Si}$, the pulse character remains single cycle but loses some high frequency components as the front overtakes the THz wave and it experiences losses in the plasma, here with a plasma frequency $\omega_p/2\pi \approx 1$~THz. \\ %= 1/2\pi \sqrt{N e^2 / m_eff / \epsilon_0 / \epsilon_r}
\indent In the sub-luminal regime, the duration of the THz field plateau is tunable with propagation length as shown in Fig.~\ref{fig:figure2}. Truncated carrier density profiles, calibrated from camera images and power measurements, are given in Fig.~\ref{fig:figure2}a with the subsequently measured THz pulses plotted in Fig.~\ref{fig:figure2}b. The positions of two sharp knife edges in the beam map directly to start $t_\text{start}$ and stop times $t_\text{end}$ of the THz pulse, corresponding to the measured pulse arising from the leading and trailing edge of the carrier density front, respectively. Sliding either knife edge by $\Delta z$ and measuring the pulse duration change $\Delta t$ allows a precise calibration of the front velocity through 
\begin{linenomath}\begin{equation}
v_f = \frac{1}{\Delta t / \Delta z + n/c},
\end{equation}\end{linenomath}
which in this case yields a sub-luminal velocity of $v_f=0.86$ $c/n_\text{Si}$. The amplitude of the plateau follows the Gaussian profile of the NIR beam due to the linearity of field emission with the local current density.\\
\begin{figure*}[t!]
\centering
\includegraphics[width=1\textwidth]{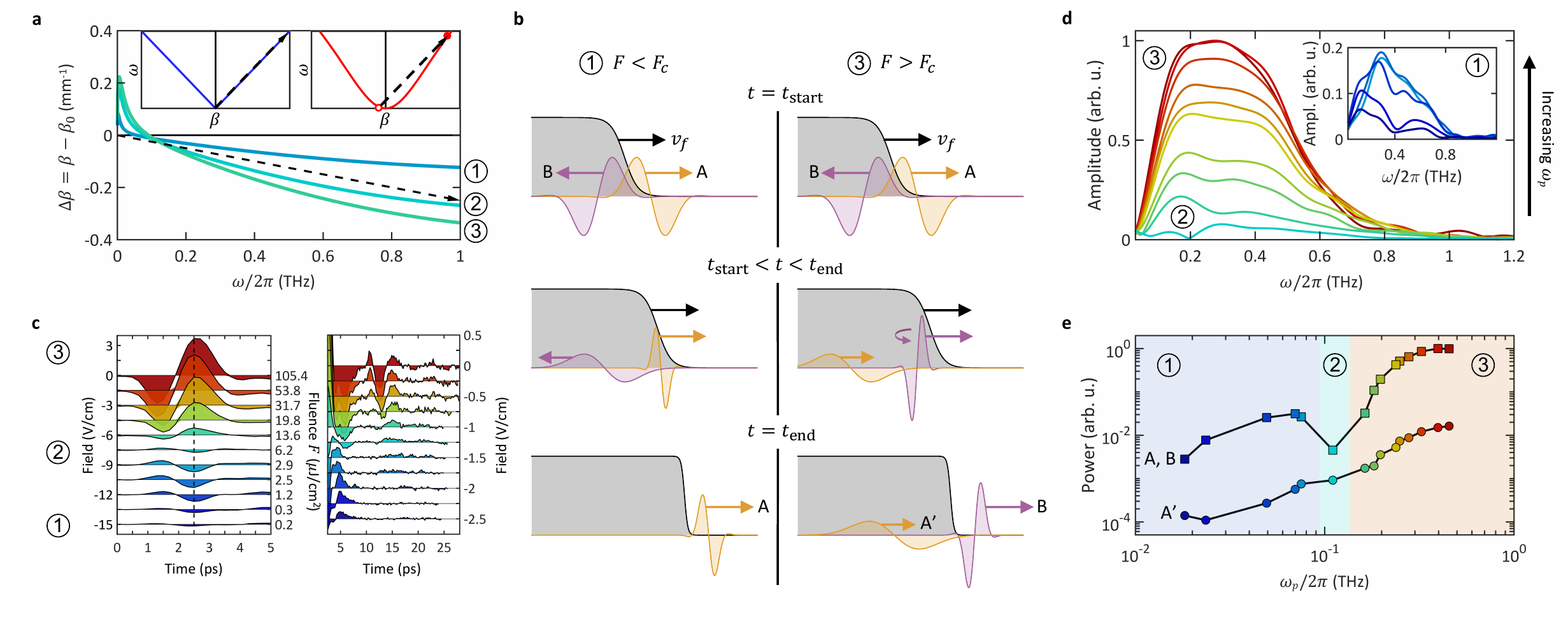}
\caption{\textbf{a}, Plasma dispersion relation $\beta$ projected onto the unexcited TEM mode wavenumber $\beta_0$ (horizontal black line) for excitation conditions: (1 - light blue line) $\omega_2/\beta_2 < v_f$, (2 - cyan line) $\omega_2/\beta_2 = v_f$ and (3 - green line) $\omega_2/\beta_2 > v_f$. The dashed line is the front velocity $v_f = 1.0035$ $c/n_\text{Si}$.  The left inset shows that only forward intra-band scattering is possible at weak excitation (blue line), whereas the right inset shows that a time-reversal scattering event ($-\beta_1\rightarrow\beta_2$, empty dot to solid circle) becomes phase matched for highly photoexcited (red line) dispersion conditions. \textbf{b}, Real space pictures that follow the evolution of initially-forward (pulse A) and initially-backward (pulse B) propagating THz waves for plasma conditions induced below and above critical fluence. \textbf{c}, Measured single cycle THz pulses generated by a slightly super-luminal front for indicated levels of pump fluence. The pulses on the left display time-reversal at $t = 2.5$ ps and, on the right, a chirped pulse builds up about 10 ps later from low-frequency waves surviving the plasma. \textbf{d}, Amplitude spectra of only the pulses displaying time-reversal in \textbf{c} for all fluences explored, categorized above critical fluence (and below in the inset). \textbf{e}, Integrated spectral power of the time-reversed THz pulses (squares) and those arising 10 ps later (circles).}
\label{fig:figure3}
\end{figure*}
\indent Two-dimensional finite-difference time-domain (2D-FDTD) simulations were performed to further understand the influence of the moving front on the THz pulse waveforms. A perfectly conducting layer was chosen to bound the domain and Schottky fields were placed on the top and bottom extending into the silicon by the Debye length (about 1 $\mu$m). The dispersion of photoexcited silicon was modelled assuming a Drude form with a scattering time of 0.1 ps and carrier densities governed by the profiles in Fig.~\ref{fig:figure2}a, reaching a maximum of $10^{17}$ cm$^{-3}$~\cite{Meng15}. The electric field of the emitted THz wave coupled to the TEM mode was extracted at a fixed position within the waveguide just outside the illuminated region and a filter function was applied in order to approximate the experimental losses from the waveguide and detection response function~\cite{Cote03}. The simulated THz pulses, shown in Fig.~\ref{fig:figure2}b agree well with the detected THz waveforms, qualitatively reproducing the variations in the plateau amplitudes.\\
\indent Setting $v_f > c/n_\text{Si}$ forces the emitted THz waves to interact with the moving front over the entire propagation distance, bringing about front-induced transitions. Deep in the super-luminal regime, the underdense plasma rapidly overtakes the THz wave and the pulse experiences both absorption and dispersion much the same as it would for transiting a static dispersive plasma~\cite{Gingras18}, as we have verified for $v_f=1.5$ $c/n_\text{Si}$ conditions. For $v_f$ very close but still exceeding $c/n_\text{Si}$, however, the front interacts with the THz emission for an extended interaction length without being consumed by the absorptive plasma. The wave therefore exists within the plasma skin depth, which is about $150$ $\mu$m at $f = 0.5$ THz and which spans a single wavelength in the photoexcited silicon, supporting single cycle pulses.\\
\indent Figure~\ref{fig:figure3} demonstrates the concept of a front-induced transition, which is explained in more detail in ref.~\cite{GaafarNatPhoton2019}. For dispersionless ($\omega_p\tau\ll1$) and underdense ($\omega_p<\omega$) plasmas where hard reflections can be ignored, the photonic band structure of the photoexcited silicon is approximately linear with a phase velocity given by
\begin{linenomath}\begin{equation}
\frac{\omega_2}{\beta_2} \approx \frac{c}{n_\text{Si}}\biggl(1+\frac{\omega_p^2\tau^2}{2}\biggr). \label{eqn:luminalfront}
\end{equation}\end{linenomath}
%We note this ignores dispersion at very low $\omega$, discussed later.
%
The two insets of Fig.~\ref{fig:figure3}a show that increasing the plasma frequency from low (blue) to high (red) adds curvature to the band, flattening at low THz frequencies. Front-induced transitions on a dispersion diagram, from an initial state $(\omega_1,\beta_1)$ to a final state $(\omega_2,\beta_2)$, must obey the phase continuity relation $\omega_1 t - z_f\beta_1 = \omega_2 t - z_f\beta_2$ where $z_f = v_f t$. The ratio of frequency change to wavenumber change is equal to the front velocity, i.e. $\Delta\omega/\Delta \beta = (\omega_2 - \omega_1)/(\beta_2-\beta_1) = v_f$. Thus the allowed transitions are tunable with the dispersion of the perturbation via the optical excitation power controlling $\omega_p$ and the pump tilt angle controlling $v_f$. The main diagram in Fig.~\ref{fig:figure3}a shows the dispersion relative to unexcited silicon for three possible perturbations with increasing $\omega_p$ labelled by 1, 2 and 3. The dispersion of the plasma is modelled by a Drude response with a fixed scattering time $\tau$ of 0.1 ps and a variable plasma frequency $\omega_p$ set by the pump fluence. In each plot, the phase continuity condition is denoted by the dashed black line where a slightly super-luminal front velocity of $v_f = 1.0035$ $c/n_\text{Si}$ has been chosen for reasons that will be explained below.\\
\indent The Schottky THz emission mechanism produces an outgoing spherical electromagnetic wave containing components travelling towards the detector (positive $\beta$) as well as those pointed backward into the plasma front (negative $\beta$). For a super-luminal front, the emission is immediately swept up by the plasma so only intra-band front-induced transitions may take place. Considering that only positive wavevectors will be detected, three scenarios are indicated in Fig.~\ref{fig:figure3}a: (1 - blue line) $\omega_2/\beta_2 < v_f$, where transitions involving solely positive wavenumbers can occur. For high fluences (3 - green line) $\omega_2/\beta_2>v_f$, transitions from backward propagating initial states to forward propagating states within the plasma become possible. In this scenario, the backward wave undergoes a time-reversal as the spatial distribution remains unchanged but the group velocity inverts sign. Higher plasma frequencies forbid intra-band scattering at lower frequencies until time-reversal scattering from $-\beta_1$ to $+\beta_2$ takes over. Here, in the intermediate regime, the special case (2 - cyan line) $\omega_2/\beta_2 = v_f$ prohibits all forward initial states from undergoing transitions. The plasma frequency that satisfies this dispersion condition allows equation~\ref{eqn:luminalfront} to be solved for the front velocity.\\
\indent Figure~\ref{fig:figure3}b illustrates in real space a picture of the pulse time-reversal we obtain through control over the plasma dispersion in the regime where $v_f$ is slightly super-luminal. At low fluence, i.e. scenario 1, the forward propagating THz component (pulse A) can be projected up to higher frequency states whereas its backwards counterpart (pulse B) is undetected, lost to the plasma. When the front stops at the edge of the photoexcited region, the blueshifted forward THz wave is ejected. In scenario 3, the optical fluence is tuned above a critical value $F_c$ for which the slope of the photoexcited band matches the front velocity. The additional curvature in the plasma dispersion prohibits fast forward waves of type A from undergoing intra-band transitions. Instead, the highly dispersive front prevents their passage and they recede back to the plasma skin depth (pulse A$'$), where they are made to co-travel with the front. Finally, in the same regime above critical fluence, the initially-backwards wave (pulse B) is able to find transitions to final states with positive $\beta$, reversing its group velocity through intra-band reflection.\\
\indent Figure~\ref{fig:figure3}c shows the influence of tuning the plasma dispersion curve on the detected THz waveforms for a slightly super-luminal velocity front. For these results we utilize a PPWG composed of ITO/Si/Au layering, where 5-10 nm of Ti is used to adhere 150 nm of Au to the bottom layer. Varying the fluence of the NIR pump pulse, we observe on the left single cycle THz pulses as expected, however for an intermediate fluence (scenario 2), we see a near extinction of the pulse followed by a $\pi$ phase inversion at higher fluences. This extinction marks the crossover from forward intra-band scattering to a time-reversal scattering event. The right plot in Fig.~\ref{fig:figure3}c highlights the development at increasing fluence of a second low-frequency THz pulse emerging roughly 10 ps later, corresponding to the type A$'$ pulses that have been held back in the low-frequency plasma skin before ejection.\\
\indent Next we take a frequency-domain view of our experiment, first by analyzing the evolution of measured type B pulses as we alter their dielectric environment. Amplitude spectra obtained from numerical fast-Fourier transform of these pulses are shown in Fig.~\ref{fig:figure3}d, revealing how the final states populate as the plasma frequency is swept through the scenarios 1, 2 and 3 described in Fig.~\ref{fig:figure3}a. A steep super-Gaussian filter has been applied in order to window the pulses. Each spectrum displays a high-frequency shoulder due to waveguide loss. As $\omega_p$ is decreased from its highest value, the spectra acquired between scenarios 3 to 2 maintain a smooth broadband shape while reducing in amplitude, consistent with front-induced transitions occurring across the entire detection bandwidth. However, in the inset, between scenarios 2 to 1 a spectral dip suddenly appears and develops near 0.4 THz. As seen in Fig.~\ref{fig:figure3}a, weak excitation produces dispersion that is predominantly linear, i.e. well-approximated by equation~\ref{eqn:luminalfront}, and we therefore expect this dip to arise from interference between co-propagating frequency states populated inside and just outside the moving plasma front, both travelling in the neighbourhood of $c/n_\text{Si}$. In Fig.~\ref{fig:figure3}e, we verify the distinct behaviour of the pulses as the plasma frequency of the excitation is tuned through the 3 identified scenarios. Type A$'$ pulses follow a continuous decrease in integrated power, whereas as type A evolves into type B a minimum occurs at scenario 2 which marks the crossover between forward intra-band scattering to time-reversing front-induced transitions followed by a resurgence and true nought.\\
\indent In summary, we have demonstrated that the SLIPSTREAM platform enables the interaction between THz light and relativistic moving fronts of photoinduced dielectric modulation. In the sub-luminal regime, the relativistic front generates a structured THz pulse displaying a quasi-static electric field profile via the extraction of energy from built-in Schottky fields at metal-semiconductor boundaries. The temporal profile of the electric field can be tailored by the intensity distribution of the optical photoexcitation. This capability allows, for example, applications where long, tunable THz pulses may act as an ultrafast optical bias~\cite{Tehini19,Wimmer14}. In addition, we showed that when the front velocity is tuned just above luminal in the silicon, the fluence dependence of the dispersion of the carrier plasma is crucial in determining the front-induced transitions that may occur. We observed time-reversal of a THz pulse as the photoinduced dispersion of the dielectric perturbation was optically tuned relative to the front velocity. This work establishes our platform as a practical avenue to explore the interaction of light with moving dielectric fronts. The ability to apply localized dielectric modulation by optical means on a silicon chip has significant implications for investigating, for example, the relativistic Doppler effect~\cite{Roskos14,Kohno20,Bakunov20}, optical horizon analogues~\cite{Faccio12} and experimentally novel effects associated with the motion of shocklike dielectric modulations travelling in photonic crystal landscapes~\cite{Reed03}.\\
%
%\acknowledgement{
\indent The authors wish to acknowledge financial support from the Natural Sciences and Engineering Research Council of Canada (NSERC), Fonds de Recherche du Qu$\acute{\text{e}}$bec Nature et Technologies (FRQNT) and the Canadian Foundation for Innovation (CFI). We also gratefully acknowledge useful discussions with E. Yablonovich. %}

\begin{nolinenumbers}
\bibliography{bibliography.bib}
\end{nolinenumbers}
\end{document}